# Polar magnetization unveiled by polarized neutron diffraction

S. W. Lovesey[1, 2]

[1]ISIS Facility, STFC, Didcot, Oxfordshire OX11 0QX, UK

[2]Diamond Light Source Ltd, Didcot, Oxfordshire OX11 0DE, UK

**Abstract** Polar magnetism is present when ions occupy sites that are not centres of inversion symmetry. Fortunately, such magnetization contributes to neutron scattering that is the bedrock of magnetic structure determinations. Experiments in which the scattered neutron polarization is analysed are not a novelty. Simulations of polarized neutron scattering amplitudes for room temperature haematite ($\alpha$-$Fe_2O_3$) demonstrate the wealth of information on offer. Two magnetic motifs distinguished by the orientation of their bulk ferromagnetism are considered. Additionally, the symmetry-inspired simulations challenge a recent claim to have determined the absolute direction of the Dzyaloshinskii-Moriya (D-M) interaction.

## I. INTRODUCTION

The well-established technique of polarized neutron diffraction (PND) has recently been promoted as a simple and direct method by which to determine the absolute direction of the Dzyaloshinskii-Moriya (D-M) interaction in weak ferromagnets [1, 2, 3, 4]. PND is a very sensitive probe of magnetic order applicable to a small class of antiferromagnets, e.g., $MnF_2$ and $NiF_2$ with Mn and Ni ions using centrosymmetric sites in a rutile-type crystal structure [5, 6]. Indeed, for the technique to be useful magnetic order must not break translation symmetry in a centrosymmetric crystal with anti-inversion absent in the magnetic crystal class. Specifically, PND is not suitable for studies of magnetic order in magneto-electric materials.

The room-temperature magnetic structure of haematite ($\alpha$-$Fe_2O_3$), so-called phase II [4, 7], meets these three material requirements. A published interpretation of PND data used for the direction of the D-M interaction is incomplete, although once recognized the shortcoming is likely a valuable asset in future use of PND [1]. The reported recovery of the D-M direction proceeds with exclusion of Dirac multipoles, e.g., an anapole or toroidal dipole depicted in Fig. 1. These polar magnetic multipoles exist when magnetic ions occupy acentric sites. Iron ions in haematite occupy sites devoid of symmetry, and Dirac multipoles surely depend on the direction of the D-M interaction. Resonant x-ray Bragg diffraction revealed Dirac multipoles in several materials, including haematite phase II [8, 9], and a direct observation of anapoles by neutron diffraction exploited polarization analysis [10]. There are no selection rules that limit the number of multipoles of given rank in the absence of any symmetry in environments inhabited by Fe ions. Whereas, in the paramagnetic phase dipoles and quadrupoles are restricted to diagonal components alone on the insistence of a triad axis of rotation symmetry.

We have calculated PND signals for haematite phase II with Dirac multipoles, and a spin-flip ratio that measures the magnetic content of a Bragg spot. Results demonstrate the wealth of knowledge available on magnetic properties, while pointing to shortcomings in the published interpretation of PND data on α-Fe$_2$O$_3$ [1]. The magnetic structure (space group) of α-Fe$_2$O$_3$ remains unsettled, and our calculations are made with two favoured candidates. Bulk ferromagnetism is confined to the basal plane in one candidate, while it is confined to a plane normal to it in a second candidate.

## II. CRYSTAL AND MAGNETIC STRUCTURES

The chemical structure of haematite is hexagonal R$\bar{3}$c (No. 167 BNS [11]), with ferric (Fe$^{3+}$, 3d$^5$) ions in sites 12c with coordinates (0, 0, 0.1447), and symmetry 3. Oxygen ions (O$^{2-}$) occupy sites 18e.

Magnetic structures use monoclinic C2/c (No. 15.85, magnetic crystal-class 2/m or C$_{2h}$) or C2′/c′ (No. 15.89, 2′/m′) with Fe ions in sites 8f devoid of symmetry. The C2/c structure was proposed by Przeniosko *et al.* [12], but no consensus has been reached. Both structures possess a centre of inversion symmetry, and permitted terms in the thermodynamic potential include H and HEE, where H and E are magnetic and electric fields. They have different piezomagnetic properties, however, with eight (C2/c) and ten (C2′/c′) elements in the corresponding tensor. Local monoclinic axes and the parent structure are related by a basis = {(− 2,− 1, 0), (0, − 1, 0), (2/3, 1/3, 1/3)} with origin (1/6,− 1/3, 1/6). Motifs of magnetic dipole moments are ($m_x$, $m_y$, $m_z$), (−$m_x$, $m_y$, −$m_z$), ($m_x$, $m_y$, $m_z$), (−$m_x$, $m_y$, −$m_z$) in C2/c, and ($m_x$, $m_y$, $m_z$), ($m_x$, −$m_y$, $m_z$), ($m_x$, $m_y$, $m_z$), ($m_x$, −$m_y$, $m_z$) in C2′/c′.

Monoclinic Miller indices are h = − (2H$_o$ + K$_o$), k = − K$_o$ and l = (1/3) (2H$_o$ + K$_o$ + L$_o$) where (H$_o$, K$_o$, L$_o$) are integer Miller indices for R$\bar{3}$c. Diffraction by Fe nuclei is forbidden with L$_o$ odd. Axes for the hexagonal structure are **a**$_h$ = $a$ (1, 0, 0), **b**$_h$ = ($a$/2) (− 1, √3, 0) and **c**$_h$ = $c$ (0, 0, 1) (lattice parameters $a$ ≈ 5.035 Å, $c$ ≈ 13.758 Å [1]). Iron multipoles are referred to orthogonal axes (ξ, η, ζ) based on (**a**$_m$*, **b**$_m$, **c**$_m$), e.g., η = **b**$_m$/|**b**$_m$|. Here, **a**$_m$ = − √3 ($a$/2) (√3, 1, 0), **b**$_m$ = − **b**$_h$ and **c**$_m$ = (1/3) (− **a**$_m$ + **c**$_h$) = ($c$/2) (t, t/√3, 2/3), with t = ($a$/$c$) and an obtuse angle β$_o$ ≈ 122.39° between **a**$_m$ and **c**$_m$. The monoclinic cell volume = $a^2c$/√3, which is 2/3 rds. of the hexagonal cell volume. Note that **a**$_m$ and **b**$_h$ = − **b**$_m$ are orthogonal vectors in the plane normal to **c**$_h$. A monoclinic reciprocal-lattice vector **a**$_m$* ∝ (√3, 1, − t 2√3) is orthogonal to **b**$_m$ and **c**$_m$. In C2′/c′ the antiferromagnetic motif of axial dipole moments uses **b**$_h$ while the ferro-component is in the plane spanned by **a**$_m$ and **c**$_m$.

Iron spherical multipoles ⟨U$^K_Q$⟩ of integer rank K possess projections Q in the interval − K ≤ Q ≤ K [13, 14]. Cartesian (x, y, z) and spherical components of a dipole **n** are related by x = (n$_{−1}$ − n$_{+1}$)/√2, y = $i$(n$_{−1}$ + n$_{+1}$)/√2 and z = n$_0$. Angular brackets ⟨ ... ⟩ denote the time-average, or expectation value, of the enclosed tensor operator. Multipoles are present in the electronic ground state.

## III. BRAGG DIFFRACTION

We adopt a unit-cell structure factor for Bragg diffraction,

$$\Psi^K_Q = [\exp(i\boldsymbol{\kappa} \cdot \mathbf{r}) \langle U^K_Q \rangle_r], \tag{1}$$

where the reflection vector $\boldsymbol{\kappa}$ is defined by $(h, k, l)$, and the implied sum in $\Psi^K_Q$ is over all Fe sites $\mathbf{r}$ in a unit cell. Environments at the four Fe sites in the monoclinic cell are related by the operations of inversion and two-fold rotation about the unique axis $\boldsymbol{\eta}$. The reflection condition $h + k$ even holds for a C-face centred cell. One finds [11],

$$\Psi^K_Q(C2/c \ \& \ C2'/c') = [1 + (-1)^{h+k}] [\alpha \langle U^K_Q \rangle + (-1)^Q \beta \langle U^K_{-Q} \rangle], \tag{2}$$

$$\alpha = [\{\exp(i\varphi) + \sigma_\pi \exp(-i\varphi)\}], \ \beta = \Phi (-1)^{K+l} \{\exp(i\varphi') + \sigma_\pi \exp(-i\varphi')\}.$$

Here, $\Phi = +1$ $(-1)$ for C2/c (C2'/c') and $\sigma_\pi = +1$ $(-1)$ for parity-even (parity-odd, Dirac) multipoles. Angles in $\alpha$ and $\beta$ are $\varphi = 2\pi(xh + yk + zl)$ and $\varphi' = 2\pi(-xh + yk - zl)$, with fractional coordinates $y = 1/4$, $x \approx 0.395$ and $z \approx 0.934$. Influential selection rules stem from identities $[\sin(\varphi) - \sin(\varphi')] = [\cos(\varphi) + \cos(\varphi')] = 0$ for $k$ odd. For, $[\sin(\varphi) - \sin(\varphi')]$ and $[\cos(\varphi) + \cos(\varphi')]$ are proportional to $\cos(\pi k/2)$, while $[\sin(\varphi) + \sin(\varphi')]$ and $[\cos(\varphi) - \cos(\varphi')]$ are proportional to $\sin(\pi k/2)$. Multipoles in Eq. (2) are time-odd (magnetic) for neutron diffraction.

It is convenient in calculations of a neutron scattering amplitude $\langle \mathbf{Q} \rangle$ to express it in terms of quantities that are even and odd functions of projections [14]. From Eq. (2) we find,

$$A^K_Q, B^K_Q = (1/2) [1 + (-1)^{h+k}] [\alpha \pm (-1)^Q \beta] [\langle U^K_Q \rangle \pm \langle U^K_{-Q} \rangle]. \tag{3}$$

The upper (lower) sign belongs to $A^K_Q$ ($B^K_Q$). Notably, $\beta$ depends on parity $\sigma_\pi$, multipole rank K, Miller index $l$ and motif signature $\Phi = \pm 1$, together with the angle $\varphi'$, while $\alpha$ depends $\sigma_\pi$ and the angle $\varphi$. In particular, Cartesian components of a dipole (K = 1) are derived from $A^1_0 \propto [\alpha + \beta] \langle U^1\zeta \rangle$, $A^1_1 \propto i[\alpha - \beta] \langle U^1\eta \rangle$ and $B^1_1 \propto [\alpha + \beta] \langle U^1\xi \rangle$.

To examine bulk properties set $h = k = l = 0$ that results in $\varphi = \varphi' = 0$. For these conditions, Dirac multipoles obey $\Psi^K_Q(C2/c \ \& \ C2'/c') = 0$. Axial multipoles $\langle \mathbf{T}^K \rangle$, with $\sigma_\pi = +1$, in the two monoclinic structures are different. Setting K = 1 in Eq. (2) it follows that $\Psi^1_0(C2/c) = 0$ together with $\Psi^1_{+1}(C2/c) \propto \langle T^1\eta \rangle$, meaning the bulk ferromagnetic moment lies along $\mathbf{b}_h$. Turning to $\Psi^1_Q(C2'/c')$, one finds $\Psi^1_0(C2'/c') \propto \langle T^1\zeta \rangle$ and $\Psi^1_{+1}(C2'/c') \propto \langle T^1\xi \rangle$, so the ferromagnetic moment lies in the plane defined by $\mathbf{a}_m$ and $\mathbf{c}_m$ that is normal to $\mathbf{b}_h$.

Most importantly, magnetic multipoles in neutron diffraction depend on the magnitude of the reflection vector, $\boldsymbol{\kappa}$. Fig. 2 shows radial integrals, also known as atomic form factors, which occur in dipoles. Dipoles $\langle \mathbf{T}^1 \rangle$ depend on standard radial integrals $\langle j_0(\kappa) \rangle$ and $\langle j_2(\kappa) \rangle$ displayed in Fig. 2, with $\langle j_0(0) \rangle = 1$ and $\langle j_2(0) \rangle = 0$. A useful result is [14],

$$\langle \mathbf{T}^1 \rangle \approx (1/3) [2\langle \mathbf{S} \rangle \langle j_0(\kappa) \rangle + \langle \mathbf{L} \rangle (\langle j_0(\kappa) \rangle + \langle j_2(\kappa) \rangle)]. \tag{4}$$

The coefficient of $\langle \mathbf{L} \rangle$ is approximate, while $\langle \mathbf{T}^1 \rangle = (1/3) \langle 2\mathbf{S} + \mathbf{L} \rangle$ for $\kappa \to 0$ is an exact result.

The magnetic polar dipole $\langle \mathbf{d} \rangle$ depends on three radial integrals displayed in Fig. 2. We use [14],

$$\langle \mathbf{d} \rangle = (1/2) [\, i(g_1) \langle \mathbf{n} \rangle + 3 (h_1) \langle \mathbf{S} \times \mathbf{n} \rangle - (j_0) \langle \mathbf{\Omega} \rangle ]. \tag{5}$$

Radial integrals $(g_1)$ and $(j_0)$ diverge in the forward direction of scattering ($\kappa \to 0$), and $(h_1)$ is also the $\kappa$-dependence of the polar spin quadrupole observed in neutron diffraction from high-$T_c$ compounds Hg1201 and YBCO [16, 17]. Dipoles $\langle \mathbf{S} \times \mathbf{n} \rangle$ and $\langle \mathbf{\Omega} \rangle = [\langle \mathbf{L} \times \mathbf{n} \rangle - \langle \mathbf{n} \times \mathbf{L} \rangle]$ are spin and orbital anapoles, and the latter (toroidal dipole) is depicted in Fig. 1. A minimal model of 3d-4p hybridization in Ref. [8] yields a guide to Dirac multipoles in resonant x-ray and neutron Bragg diffraction. The axial dipole is suitably small and spin-only in the model. Corresponding Dirac multipoles that contribute to resonant x-ray and neutron diffraction are found to be similar. This finding bolsters the relevance of the reported observation of Dirac multipoles in haematite by x-rays [9] in a proper analysis of neutron diffraction by haematite. An important difference between amplitudes for resonant x-ray and neutron diffraction is the absence of radial integrals in the former. Notably, all radial integrals in the polar dipole displayed in Fig. 2 are of a similar size at the Bragg spot $(-7, 1, 2)_m$ where the reflection vector $\kappa \approx 5.22$ Å$^{-1}$.

## IV. POLARIZED NEUTRON DIFFRACTION

A polarized neutron diffraction (PND) signal $\Delta = \{\mathbf{P} \cdot [\mathbf{e} \times (\langle \mathbf{Q} \rangle \times \mathbf{e})]\}$, where $\mathbf{P}$ is polarization of the primary neutrons and $\mathbf{e}$ is a unit reflection vector $\mathbf{e} = \boldsymbol{\kappa}/\kappa$. In the case of haematite phase II, the signal is in phase with nuclear diffraction by ligand ions. This is not the case for eskolaite (chromium sesquioxide, $Cr_2O_3$), for example, a paradigm for the linear magneto-electric effect (magnetic crystal-class $\bar{3}'m'$). Intensity of a Bragg spot $= |\langle \mathbf{Q} \rangle - \mathbf{e}\, (\langle \mathbf{Q} \rangle \cdot \mathbf{e})|^2$. Thoma *et al*. chose $\mathbf{P}$ parallel to $\mathbf{a}_h + 2\mathbf{b}_h = a\, (0, \sqrt{3}, 0)$ [1]. The authors deduce the absolute direction of the D-M interaction from the reflection indexed by $(4, -1, -1)_h \equiv (-7, 1, 2)_m$. It is a strong out-of-plane Bragg spot for which the Fe nuclear structure factor is zero. Referring to radial integrals displayed in Fig. 2, the result $\kappa = (2\pi/a\sqrt{3})\, [3H_o^2 + (H_o + 2K_o)^2 + 3(t\, L_o)^2]^{1/2}$ yields $\kappa \approx 5.22$ Å$^{-1}$ for $(4, -1, -1)_h$.

The complete PND signal to be confronted with a Bragg diffraction pattern $\Delta = (\Delta^{(+)} + \Delta^{(-)})$ with axial and Dirac contributions labelled by our parity signature $\sigma_\pi$ ($\pm 1$). For axial multipoles,

$$\Delta^{(+)} = (Z^2/6) \{\sin(\beta_o) \langle Q_\xi \rangle^{(+)} [k\, (h - k) + 3t^2\, (k - l)\, (h + 3l)]$$
$$- \langle Q_\eta \rangle^{(+)} (1/\sqrt{3}) [h\, (h - k) + 3t^2\, (h + 3l)^2] \tag{6}$$
$$- \cos(\beta_o) \langle Q_\zeta \rangle^{(+)} (t/\sqrt{3}) [(1/3)\, h^2 + k^2 + l\, (h + 3k) + t^2\, (h + 3l)^2]\}.$$

The diffraction amplitude $\langle \mathbf{Q} \rangle = (\langle Q_\xi \rangle \boldsymbol{\xi}, \langle Q_\eta \rangle \boldsymbol{\eta}, \langle Q_\zeta \rangle \boldsymbol{\zeta})$. Eq. (6) is a quadratic function of Miller indices and, thus, identical for $(h, k, l)$ and $(-h, -k, -l)$. Likewise, $\Delta^{(-)}$ derived from Dirac multipoles to be considered. In Eq. (6),

$$R = [1 + 3t^2]^{1/2}, \quad Z^2 \left[ (1/3) h^2 + k^2 + t^2 (h + 3l)^2 \right] = 3, \qquad (7)$$

with the obtuse angle $\cos(\beta_o) = -t\sqrt{3}/R$ and $\sin(\beta_o) = (1/R)$. By way of an interesting example, the reflection vector $(2, -1, 3)_h \equiv (-3, 1, 2)_m$ is parallel to $\mathbf{a}_h$ and normal to $\mathbf{P}$. In this case, $Z^{-2} = [(4/3) + 3t^2]$ and,

$$\Delta^{(+)} = -(1/2) \{ \sin(\beta_o) \langle Q_\xi \rangle^{(+)} + \sqrt{3} \langle Q_\eta \rangle^{(+)} + \cos(\beta_o) \langle Q_\zeta \rangle^{(+)} \}. \quad (-3, 1, 2)_m \qquad (8)$$

Evaluated at the level of dipoles,

$$\langle Q_\xi \rangle^{(+)} \approx (3/2) [\alpha + \beta] \langle T^1_\xi \rangle, \quad \langle Q_\eta \rangle^{(+)} \approx (3/2) [\alpha - \beta] \langle T^1_\eta \rangle,$$

$$\langle Q_\zeta \rangle^{(+)} \approx (3/2) [\alpha + \beta] \langle T^1_\zeta \rangle. \qquad (9)$$

Higher order axial multipoles in $\langle \mathbf{Q} \rangle^{(+)}$ include a quadrupole proportional to $\langle j_2(\kappa) \rangle$ that is allowed by an admixture of manifolds in the Fe electronic ground state [14, 18].

A general expression for the PND signal derived from Dirac multipoles is cumbersome, unlike the axial case Eq. (6) that is correct for all allowed multipoles. Absence of selection rules on projections in multipoles is largely responsible for cumbersome expressions. At a first level that uses anapoles $(K = 1)$,

$$\Delta^{(-)} \approx iZ/(2\sqrt{3}) \{ \cos(\beta_o) \langle Q_\xi \rangle^{(-)} (k + 3l)$$

$$+ \langle Q_\eta \rangle^{(-)} t (h + 3l) + \sin(\beta_o) \langle Q_\zeta \rangle^{(-)} [h - k + 3t^2(h + 3l)] \}. \qquad (10)$$

In this result,

$$\langle Q_\xi \rangle^{(-)} = (1/2) [\alpha + \beta] \langle d_\xi \rangle, \quad \langle Q_\eta \rangle^{(-)} = (1/2) [\alpha - \beta] \langle d_\eta \rangle,$$

$$\langle Q_\zeta \rangle^{(-)} = [\alpha + \beta] \langle d_\zeta \rangle, \qquad (11)$$

with $\langle \mathbf{d} \rangle$ defined in Eq. (5). Using $\sigma_\pi = -1$, the functions $\alpha$ and $\beta$ in Eq. (2) are proportional to $i\sin(\varphi)$ and $i\sin(\varphi')$, respectively, that have opposite signs for $(h, k, l)$ and $(-h, -k, -l)$. Beyond the approximate result Eq. (10), we have mentioned the important role of the polar spin quadrupole in an analysis of diffraction patterns for Hg1201 and YBCO, for example, and it is proportional to the radial integral $(h_1)$ contained in Fig. 2. The result Eq. (10) for $(-3, 1, 2)_m$ does not undergo a great simplification unlike Eq. (8).

Applications of identities $[\sin(\varphi) - \sin(\varphi')] = [\cos(\varphi) + \cos(\varphi')] = 0$ for Miller index $k$ odd, $l$ even and $K = 1$ in Eqs. (9) and (11) are encapsulated by $\langle \mathbf{Q}(C2/c) \rangle = (\langle \mathbf{Q}_\xi \rangle^{(+)}, \langle \mathbf{Q}_\eta \rangle^{(-)}, \langle \mathbf{Q}_\zeta \rangle^{(+)})$ and $\langle \mathbf{Q}(C2'/c') \rangle = (\langle \mathbf{Q}_\xi \rangle^{(-)}, \langle \mathbf{Q}_\eta \rangle^{(+)}, \langle \mathbf{Q}_\zeta \rangle^{(-)})$. These relations are particular examples of general identities valid for Miller index $k$ odd, namely, $[\alpha \pm \beta]^{(+)} = [\alpha - (\pm)\beta]^{(-)} = 0$ with $\Phi (-1)^{K+l} = \pm 1$.

A so-called spin-flip intensity is the fraction of neutrons that participate in events that change (flip) the neutron spin orientation [16, 17]. For a collinear magnetic motif and complete polarization, $P^2 = 1$, the fraction,

$$\text{SF} = \{|\langle \mathbf{Q}_\perp \rangle|^2 - |\mathbf{P} \cdot \langle \mathbf{Q}_\perp \rangle|^2\}, \quad (12)$$

with $\langle \mathbf{Q}_\perp \rangle = [\mathbf{e} \times (\langle \mathbf{Q} \rangle \times \mathbf{e})]$ is a measure of the magnetic content of a Bragg spot. For the special case $\mathbf{P} \cdot \mathbf{e} = 0$, Eq. (12) reduces to $\text{SF} = \{|\langle \mathbf{Q} \rangle|^2 - |\mathbf{e} \cdot \langle \mathbf{Q} \rangle|^2 - |\mathbf{P} \cdot \langle \mathbf{Q} \rangle|^2\}$. Applied to orthogonal $\mathbf{e} \propto \mathbf{a}_h$ and $\mathbf{P} \propto \mathbf{b}_h^*$, the polarization used in Ref. [1], the spin-flip intensity measures the component of $\langle \mathbf{Q} \rangle = (\langle \mathbf{Q} \rangle^{(+)} + \langle \mathbf{Q} \rangle^{(-)})$ parallel to the crystal $\mathbf{c}$ axis, i.e., $\text{SF} = |\cos(\beta_o) \langle Q_\xi \rangle - \sin(\beta_o) \langle Q_\zeta \rangle|^2$. In consequence, SF at the Bragg spot $(-3, 1, 2)_m$ measures diffraction by axial dipoles in C2/c and anapoles in C2′/c′, respectively.

## V. DISCUSSION

In summary, our calculations of neutron diffraction amplitudes for room-temperature haematite ($\alpha$-$Fe_2O_3$, phase II) demonstrate a wealth of information in a Bragg diffraction pattern. Calculations encompass two candidate magnetic structures and axial and polar (Dirac) multipoles. The latter exist in phase II according to published resonant x-ray diffraction data [9], and anapoles are known to deflect neutrons [10]. The open question posed by the magnetic structure is not adequately addressed, and Dirac multipoles find no place in a recent account of a polarized neutron diffraction (PND) study of haematite [1].

In more detail, axial magnetic signatures to be confronted with a measurement of the Bragg spot $(-7, 1, 2)_m$ in Ref. [1] are shown by us to be $\Delta^{(+)} \propto [-\langle Q_\xi \rangle^{(+)} + 0.74 \langle Q_\zeta \rangle^{(+)}]$ or $\Delta^{(+)} \propto \langle Q_\eta \rangle^{(+)}$ for candidate magnetic space groups C2/c and C2′/c′, respectively. Here, $\langle \mathbf{Q} \rangle^{(+)}$ is the axial neutron scattering amplitude and ($\xi, \eta, \zeta$) Cartesian vectors in the monoclinic magnetic structure with unique axis $\eta$. The mentioned expressions are correct at the level of dipoles (K = 1), results in Eq. (9) apply, and $\langle \mathbf{Q} \rangle^{(+)}$ is proportional to the axial magnetic moment $\langle 2\mathbf{S} + \mathbf{L} \rangle$ in the forward direction of scattering. With regard to higher order axial multipoles, quadrupoles might be influential for two reasons. First, such quadrupoles are proportional to the radial integral $\langle j_2(\kappa) \rangle$. Reference to results for radial integrals in Fig. 2 shows that $\langle j_0(\kappa) \rangle$ and $\langle j_2(\kappa) \rangle$ that together make up $\langle \mathbf{Q} \rangle^{(+)}$ are almost equal in magnitude at the Bragg spot $(-7, 1, 2)_m$ where the reflection vector $\kappa \approx 5.22$ Å$^{-1}$. Secondly, absence of symmetry of Fe sites allows for significant mixing of atomic manifolds and an axial quadrupole, or an inextricable knot of spin and space using the spin anapole represented by the operator equivalent $\mathbf{n}(\mathbf{S} \times \mathbf{n})$ [14, 18]. Eqs. (10) and (11) for anapoles reveal PND signals $\Delta^{(-)} \propto [\langle Q_\xi \rangle^{(-)} + 1.89 \langle Q_\zeta \rangle^{(-)}]$ and $\Delta^{(-)} \propto \langle Q_\eta \rangle^{(-)}$ for space groups C2′/c′ and C2/c, respectively, for the Bragg spot $(-7, 1, 2)_m$.

With future neutron diffraction experiments on room-temperature haematite in mind, we report the spin-flip intensity for the convenient Bragg spot $(-3, 1, 2)_m$, $\kappa \approx 2.85$ Å$^{-1}$. The ratio measures the magnetic content of a Bragg spot, and the technique has been used extensively to study ceramic superconductors, e.g., Hg1201 and YBCO [19]. Axial dipoles in C2/c and anapoles in C2′/c′ are detected, respectively, for the chosen conditions of orthogonal neutron polarization and reflection vector.

**ACKNOWLEDGEMENTS** Dr D. D. Khalyavin provided details of the magnetic space groups. Dr V. Scagnoli is the author of Fig. 1. Professor G. van der Laan calculated radial integrals that appear in Fig. 2 and created the figure.

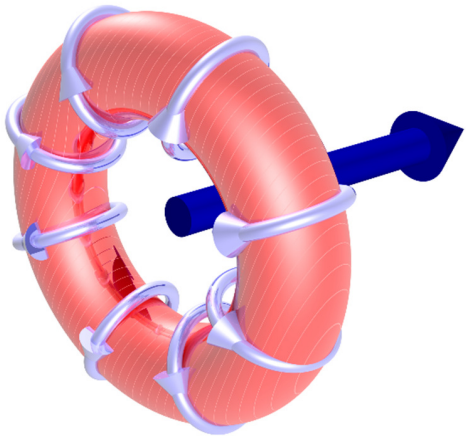

**FIG**. 1. Depiction of an orbital anapole (toroidal dipole); author V. Scagnoli.

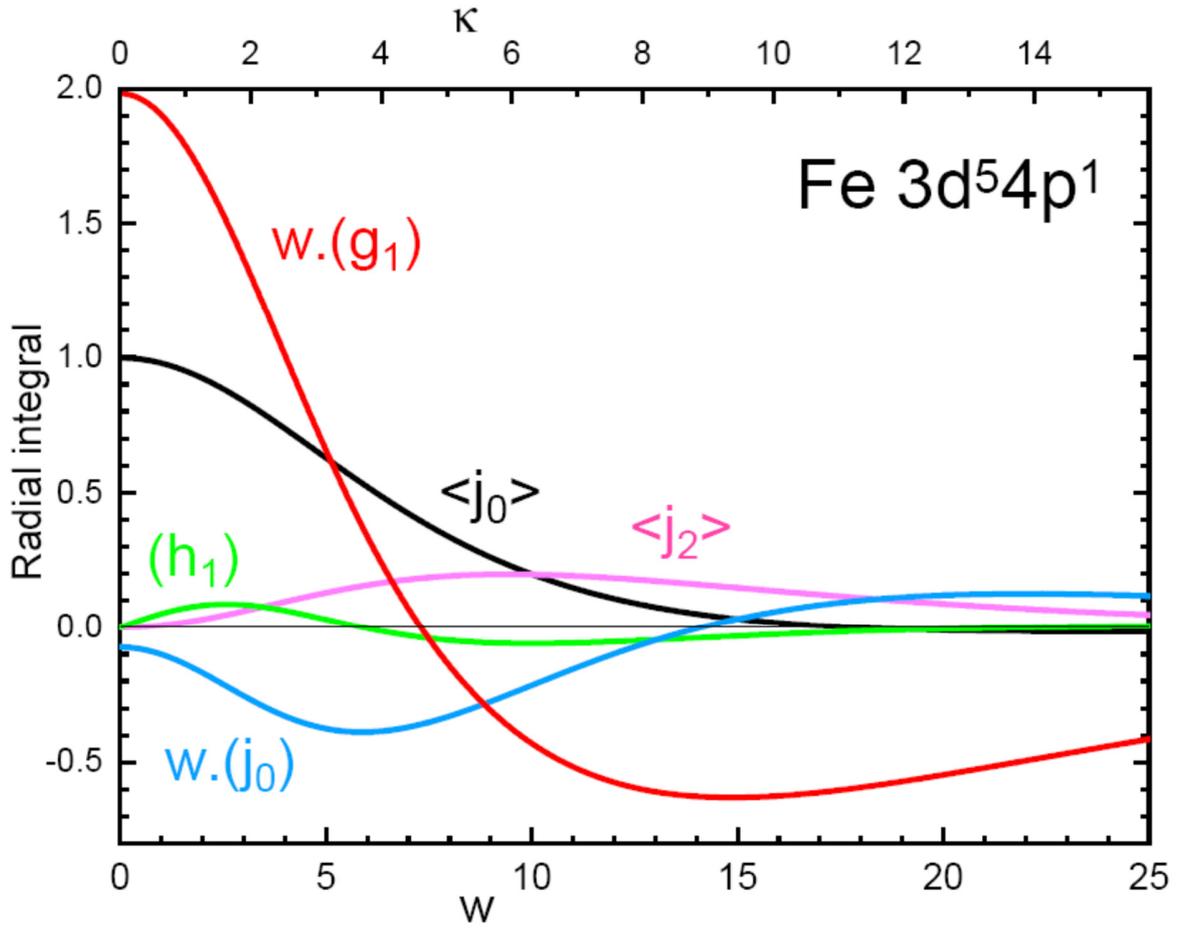

**FIG**. 2. Radial integrals for the ferric ion $Fe^{3+}(3d^5)$ displayed as a function of the magnitude of the reflection vector $\kappa = 4\pi s$ with $s = \sin(\theta)/\lambda$ (Å$^{-1}$), Bragg angle $\theta$ and neutron wavelength $\lambda$. Also, $w = 3a_o\kappa$ where $a_o$ is the Bohr radius. Blue and purple lines depict standard radial integrals $\langle j_0(\kappa)\rangle$ and $\langle j_2(\kappa)\rangle$ that occur in the axial dipole Eq. (4). Red, green and blue curves radial integrals in the polar dipole Eq. (5). Two integrals $(g_1)$ and $(j_0)$ diverge in the forward direction of scattering and $w(g_1)$ and $w(j_0)$ are displayed. Calculations using Cowan's atomic code [15] and figure by G. van der Laan.